\documentclass[12pt]{article}
\usepackage{epsf,epsfig,amsmath,amssymb,bm}
\usepackage{graphicx}
\usepackage{amssymb}

\begin{document}
\hfill{RUB-TPII-07/08}
\bigskip

\begin{center}
{\Large\bf \boldmath Renormalization-group approach to
                     transverse-momentum dependent parton
                     distribution functions in QCD\footnote{Invited
                     plenary talk presented by the first author at
                     XIII International Conference on 
                     Selected Problems of Modern Theoretical Physics,
                     Dubna, Russia, June  23-27, 2008.}} 

\vspace*{6mm}
{N.~G.~Stefanis$^a$ and I.~O.~Cherednikov$^b$ }\\      
{\small \it $^a$ Institut f\"{u}r Theoretische Physik II,
                 Ruhr-Universit\"{a}t Bochum,
                 D-44780 Bochum, Germany       
                 E-mail: stefanis@tp2.ruhr-uni-bochum.de \\
            $^b$ Bogoliubov Laboratory of Theoretical Physics,
                 JINR, 141980 Dubna, Russia \\
                 E-mail: igorch@theor.jinr.ru}
\end{center}

\vspace*{6mm}

\begin{abstract}
We discuss the renormalization of gauge-invariant transverse-momentum
dependent (TMD), i.e., unintegrated, parton distribution functions (PDFs)
and carry out the calculation of their anomalous dimension at one loop.
We show that in the light-cone gauge, TMD PDFs contain UV divergences
that may be attributed to the renormalization effect on a cusp-like
junction point of the gauge contours at infinity.
In order to eliminate the anomalous dimension ensuing from this
cusp, we propose to use in the definition of the TMD PDFs, a soft
counter term in terms of a path-ordered phase factor along a
particular cusped contour extending to transverse light-cone
infinity and comprising light-like and transverse segments.
We argue that this additional factor is analogous to the ``intrinsic''
Coulomb phase factor found before in QED.
\end{abstract}

\vspace*{6mm}

\section{Introduction}
\label{sec:intro}
Parton distribution functions encode the nonperturbative
hadronization dynamics at the amplitude level and are, therefore, of
fundamental importance in QCD calculations and phenomenological
applications (see \cite{CSS89} for a review).
While integrated PDFs can be given an unambiguous gauge-invariant
definition in terms of Wilson-line operators (gauge links) \cite{CS82},
the analogous definition for unintegrated, i.e., transverse-momentum
dependent, PDFs may depend more critically on the details of the gauge
contour.
As a result, the renormalization of TMD PDFs is a more demanding
task to which the present report is devoted.

Indeed, in order to satisfy factorization, one cannot restore gauge
invariance in TMD PDFs by inserting a purely light-like Wilson line
joining the quark and antiquark field points directly \cite{Col03}.
The reason is that the gluons emitted from the struck quark along
the $x^-$ direction have rapidities that cannot match those of the
spectator quarks moving along the $x^+$ direction.
Consequently, one is forced to employ a gauge contour that
comprises segments going off the light cone and joins the quark field
points through infinity.
Recall in this context that the gauge link resums the contributions
due to collinear and transverse gluons between the struck quark and
the spectator remnants (see, e.g., \cite{ER80RNC}).

\begin{figure}[h]
\centerline{\includegraphics[width=7cm,scale=0.5,angle=90]{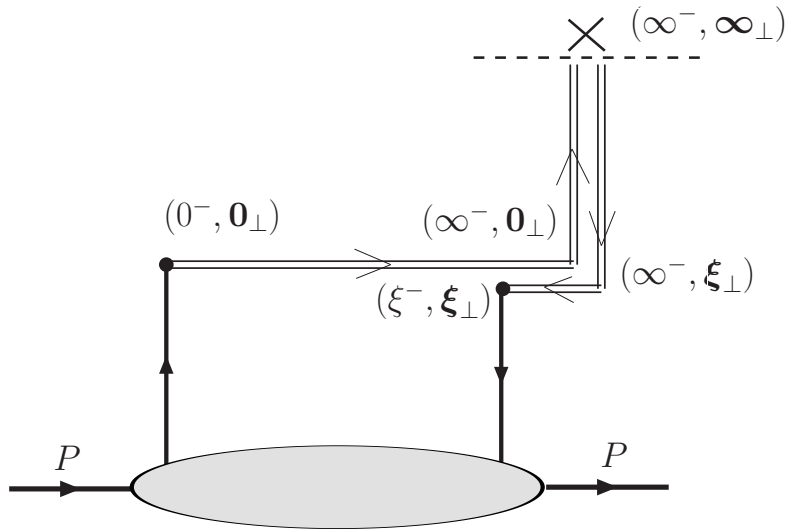}}
\caption{TMD PDF (shaded oval) in coordinate space.
         Double lines denote lightlike and transverse gauge
         links, connecting the quark field points
         $(0^{-},{\bf 0}_{\perp})$ and
         $(\xi^{-},\mbox{\boldmath$\xi_\perp$})$,
         via a composite contour through light-cone infinity
         $(\infty^-,\mbox{\boldmath$\infty_\perp$})$.
\label{fig:generic-TMD-PDF}
         }
\end{figure}

Quite recently, Belitsky, Ji, and Yuan \cite{BJY02} (see also
\cite{BHS02,JY02,BMP03}) have shown that
in the light-cone gauge $A^+=0$, one has to include in the definition
of TMD PDFs transverse gauge links at light-cone infinity---as
illustrated in Fig.\ \ref{fig:generic-TMD-PDF}.
These transverse gauge links cancel when the integration over
$\mathbf{k}_\perp$ is performed, so that one recovers the correct
integrated PDF.
Moreover, it was advocated in \cite{BJY02} that adopting the
advanced boundary condition (see below), the transverse field
$\mathbf{A}_\perp$ vanishes at $\xi^- =\infty$ reducing the
transverse gauge link to unity.
For that particular boundary condition, the light-cone gauge
(one-loop) calculation reproduces the Feynman-gauge PDF.

In these investigations it was tacitly assumed that the
lightlike-transverse composite contour going through infinity,
illustrated in Fig.\ \ref{fig:generic-TMD-PDF}, is everywhere smooth.
However, we have shown in \cite{CS07,CS08} by carrying out a one-loop
calculation of the gluon radiative corrections to the unpolarized TMD
PDF of a quark in a quark in the light-cone gauge that there are UV
divergences which are neither related to the quark self energy nor
are they caused by the endpoints of the line integral along the gauge
contour---as one finds for the direct contour (the ``connector''
\cite{CD80,Ste83}).
Instead, the origin of these extra UV divergences can be attributed
to a cusp obstruction (denoted by the symbol $\times$ in Fig.\
\ref{fig:generic-TMD-PDF}) in the split gauge contour at transverse
light-cone infinity.
The concomitant anomalous dimension after renormalization is a local
footprint of the cusp and peculiar to the split contour.
It turns out to coincide with the leading-order (LO) cusp anomalous
dimension \cite{KR87}.\footnote{It remains to be proved that this
coincidence persists at the two-loop order and beyond.}
The appearance of this extra anomalous dimension necessitates a
modification of the definition of the TMD PDF in order to dispense
with it.
As pointed out in \cite{CS07}, and further outlined in full detail in
\cite{CS08}, this can be achieved by including a path-ordered soft
factor, in the sense of Collins and Hautmann \cite{CH00}, to be
evaluated along a specific gauge contour off-the-light cone
(see next section).
Having described the cornerstones of our approach, let us now have a
closer look to its mathematical details.

\section{One-loop radiative corrections to gauge-invariant \\
         TMD PDFs}
\label{sec:TMD-PDF}

Taking into account the findings of \cite{BJY02}, the strictly
gauge-invariant operator definition of the TMD distribution of a
quark with momentum
$k_\mu = (k^+, k^-, \mathbf{k}_\perp)$
in a quark with momentum $p_\mu = (p^+, p^-, \mathbf{0}_\perp)$, with
non-lightlike Wilson lines to light-cone infinity included, reads
\begin{equation}
\begin{split}
   f_{q/q}(x, \mbox{\boldmath$k_\perp$})
 ={} &
   \frac{1}{2} \!
   \int \! \frac{d\xi^- d^2
   \mbox{\boldmath$\xi_\perp$}}{2\pi (2\pi)^2}\,
   \exp\left(- i k^+ \xi^- \!
   +\!  i \mbox{\boldmath$k_\perp$}
\cdot \mbox{\boldmath$\xi_\perp$}\right)
   \left\langle  q(p) |\bar \psi (\xi^-, \mathbf{\xi}_\perp)
   [\xi^-, \mbox{\boldmath$\xi_\perp$};
   \infty^-, \mbox{\boldmath$\xi_\perp$}]^\dagger \right.\\
& \times [\infty^-, \mbox{\boldmath$\xi_\perp$};
   \left. \infty^-, \mbox{\boldmath$\infty_\perp$}]^\dagger
   \gamma^+[\infty^-, \mbox{\boldmath$\infty_\perp$};
   \infty^-, \mbox{\boldmath$0_\perp$}]
   [\infty^-, \mbox{\boldmath$0_\perp$};0^-, \mbox{\boldmath$0_\perp$}]
   \right. \\
& \times \left. \psi (0^-,\mbox{\boldmath$0_\perp$}) |q(p)\right\rangle \
   |_{\xi^+ =0}\ . \vspace{-1cm}
\label{eq:TMD-PDF}
\end{split}
\end{equation}
Here the gauge links, in the lightlike and the transverse direction,
respectively, are defined by the following path-ordered exponentials
\begin{equation}
\begin{split}
[ \infty^-, \mbox{\boldmath$z_\perp$}; z^-, \mbox{\boldmath$z_\perp$}]
\equiv {} &
 {\cal P} \exp \left[
                     i g \int_0^\infty d\tau \ n_{\mu}^- \
                      A_{a}^{\mu}t^{a} (z + n^- \tau)
               \right] \\
[ \infty^-, \mbox{\boldmath$\infty_\perp$};
 \infty^-, \mbox{\boldmath$\xi_\perp$}]
\equiv {} &
 {\cal P} \exp \left[
                     i g \int_0^\infty d\tau \ \mbox{\boldmath$l$}
                     \cdot \mbox{\boldmath$A$}_{a} t^{a}
                     (\mbox{\boldmath$\xi_\perp$}
                     + \mbox{\boldmath$l$}\tau)
               \right] \, ,
\label{eq:gauge-links}
\end{split}
\end{equation}
where the two-dimensional vector $\mathbf{l}$ is arbitrary with no
influence on the (local) anomalous dimensions we are interested in.
\begin{figure}[hb]
\centering
\includegraphics[scale=0.65,angle=90]{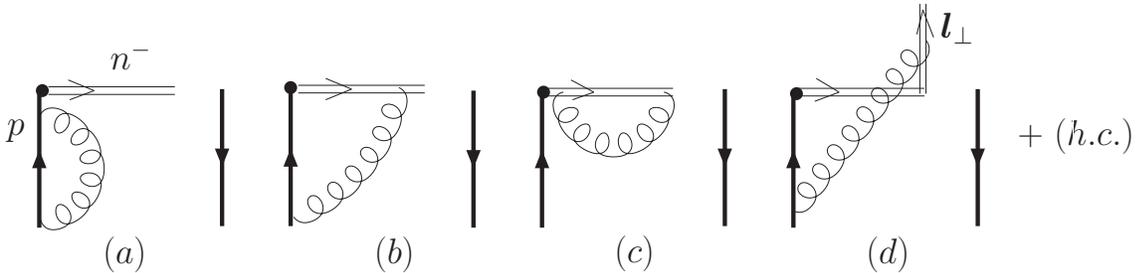}~~
\caption{One-loop radiative corrections (curly lines) contributing
         UV-divergences to $f_{q/q}(x, \mbox{\boldmath$k_\perp$})$ in a
         general covariant gauge.
         Double lines denote lightlike and transverse gauge links.
         Diagrams (b) and (c) are absent in the light-cone gauge, while
         the Hermitian conjugate (``mirror'') diagrams (not shown)
         are abbreviated by $(h.c)$.
\label{fig:se_gluon}}
\end{figure}

\begin{figure}[t]
\centerline{%
\includegraphics[width=0.28\textwidth]{%
                 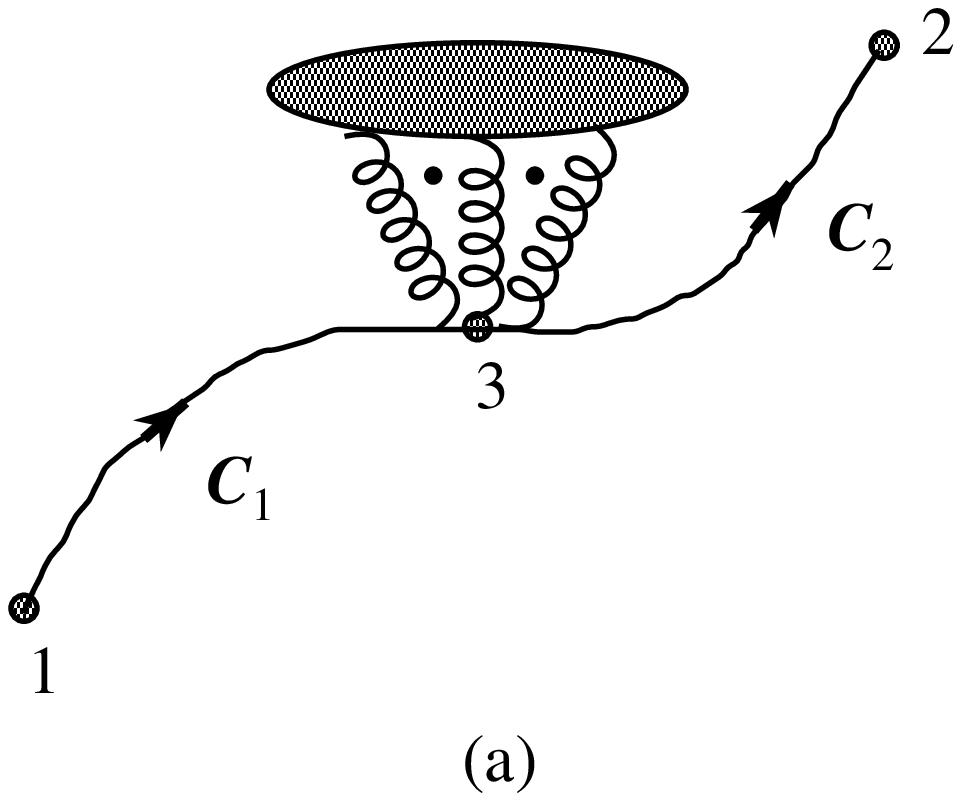}
~~~~~~~~~~~~~~~~~~~~~~%
\includegraphics[width=0.26\textwidth]{%
                 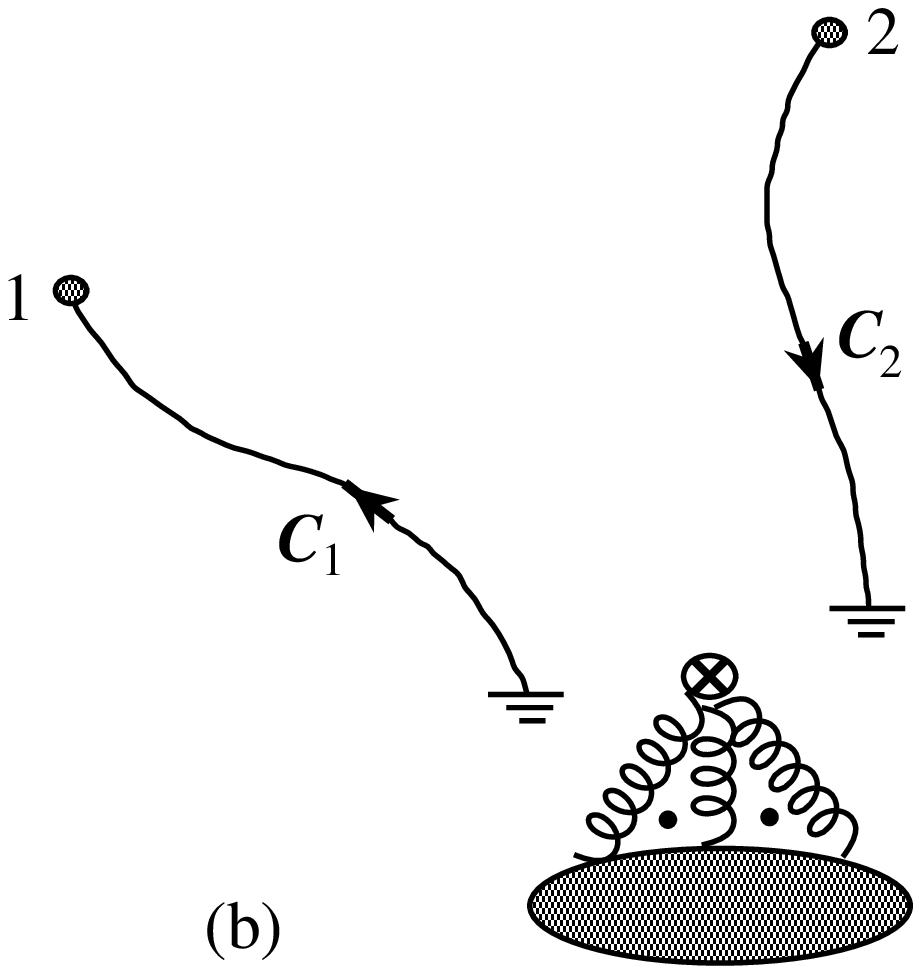}}
\caption{Renormalization effect on the junction point due to gluon
         corrections (illustrated by a shaded oval with gluon lines
         attached to it) for (a) two smoothly joined gauge contours
         $\mathcal{C}_1$ and $\mathcal{C}_2$ at point 3 and (b) the
         same for two contours joined by a cusp (indicated by the
         symbol $\otimes$) at infinite transverse distance (marked
         by the earth symbol) off the light cone.
         All contours shown are assumed to be arbitrary non-lightlike
         paths in Minkowski space.
\label{fig:cont-fact}}
\end{figure}

Employing the light-cone gauge
$
 A^+
=
 (A \cdot n^-) = 0 \ , \ {(n^-)}^2= 0
$,
we calculated in \cite{CS07,CS08} gluon radiative corrections to
$f_{q/q}(x, \mbox{\boldmath$k_\perp$})$ at the one-loop level
and identified its UV divergences (see Fig.\ \ref{fig:se_gluon}).
We found that those contributions stemming from the interactions with
the gluon field of the transverse gauge link cancel all terms that
bear a dependence on the pole prescription applied to regularize
the light-cone singularities of the gluon propagator.
In the intermediate steps of the calculation the light-cone
singularities of the gluon propagator
\begin{equation}
   D_{\mu\nu}^{\rm LC} (q)
=
   \frac{-i}{q^2 - \lambda^2 + i0} \Big( g_{\mu\nu}
  -\frac{q_\mu n^-_\nu + q_\nu n^-_\mu}{[q^+]}\Big)
\end{equation}
are taken into account by means of the term $1/[q^+]$ subject to boundary
conditions on the gauge potential.
In the present work we apply the following regularization prescriptions
to the pole at $q^+$ \cite{BJY02}:
\begin{equation}
  \frac{1}{[q^+]}\Bigg|_{\rm Ret/Adv}
=
  \frac{1}{q^+ \pm i \eta} \ \ \ , \ \ \
  \frac{1}{[q^+]}\Bigg|_{\rm PV}
=
  \frac{1}{2} \left[ \frac{1}{q^+ + i \eta}
  + \frac{1}{q^+ - i \eta} \right] \ ,
\label{eq:pole-prescription}
\end{equation}
where $\eta$ is a mass-scale parameter kept small but finite.
The total UV-divergent contribution is obtained by including also the
Hermitian conjugate contributions of diagrams (a) and (d) in Fig.\
\ref{fig:se_gluon}.
Then, we obtain
\begin{eqnarray}
  {\Sigma}^{(a+d)}_{\rm UV} (p, \mu, \alpha_s ; \epsilon)
& = &
  - \frac{\alpha_s}{\pi}\ C_{\rm F} \frac{1}{\epsilon}
    \left[
          \frac{1}{4}- \frac{\gamma^{+} \hat p}{2 p^{+}}
    \left(1 +  \ln \frac{\eta}{p^{+}} - \frac{i\pi}{2}
          - i \pi \ C_{\infty} + i \pi C_{\infty}
  \right)
    \right]
\nonumber \\
& = &
  - \frac{\alpha_s}{\pi}\ C_{\rm F} \frac{1}{\epsilon}
    \left[1 - \frac{\gamma^{+} \hat p}{2 p^{+}}
    \left(1 + \ln \frac{\eta}{p^{+}} - \frac{i\pi}{2} \right)
    \right]
    \ ,
\label{eq:s_tot}
\end{eqnarray}
where $C_{\rm F}=(N_{c}^{2}-1)/(2N_{c})=4/3$ and the parameter
$C_{\infty}$ encodes the adopted pole prescription
(cf.\ Eq.\ (\ref{eq:pole-prescription})).
This expression can be further simplified using
$$
  \frac{ \gamma^+ \hat p \gamma^+}{2 p^+} = \gamma^+
$$
and recalling that the mirror counterparts of the evaluated diagrams
yield complex-conju\-gated contributions.
As a result, the imaginary terms in Eq.\ (\ref{eq:s_tot}) mutually cancel
and one is left with
\begin{equation}
  \Sigma_{\rm UV}^{\rm (a+d)}(\alpha_s, \epsilon)
=
   2\frac{\alpha_s}{\pi}C_{\rm F} \left[ \frac{1}{\epsilon}
   \left( \frac{3}{4}
  + \ln \frac{\eta}{p^+} \right) - \gamma_E + \ln 4\pi \right]\, .
\label{eq:gamma_1}
\end{equation}
The key contribution here is the term $\sim \ln \frac{\eta}{p^+}$
which gives rise to the one-loop anomalous dimension in the light-cone
(LC) gauge
$
 \left(\gamma
=
  \frac{\mu}{2}
  \frac{1}{Z}
  \frac{\partial\alpha_s}{\partial\mu}
  \frac{\partial Z}{\partial\alpha_s}
\right):
$
\begin{equation}
  \gamma_{\rm 1-loop}^{\rm LC}
=
  \frac{\alpha_s}{\pi}C_{\rm F}\Bigg( \frac{3}{4}
  + \ln \frac{\eta}{p^+} \Bigg)
=
  \gamma_{\rm smooth} - \delta \gamma \ .
\label{eq:gamma_2}
\end{equation}
Here $\gamma_{\rm smooth}$ is the anomalous dimension one would
obtain in a covariant gauge, or, equivalently, the anomalous
dimension associated with a direct smooth contour between the quark
fields (i.e., with the connector correction).
The term $\delta \gamma$ is the anomalous-dimensions defect entailed
by the cusp, we have to compensate in order to recover the same
expression as in a covariant gauge according to the factorization
proof.
Consistent with this finding, one has to modify the multiplication
rule for gauge links (or, equivalently, the way of decomposing gauge
contours) \cite{CS08}:
\begin{equation}
  \gamma_{\mathcal{C}}
=
  \gamma_{\mathcal{C}_{1}^{\infty} \cup\, \mathcal{C}_{2}^{\infty}
           }
  +\gamma_{\rm cusp}
  ~~ \Longleftrightarrow ~~
  [2,1|\mathcal{C}]
=
  [2,\infty|\mathcal{C}_{2}^{\infty}]^{\dag}
  [\infty,1|\mathcal{C}_{1}^{\infty}]
  {\rm e}^{i \Phi_{\rm cusp}
          } \, .
\label{eq:mod-mult-rule}
\end{equation}
The graphics at right of Fig.\ \ref{fig:cont-fact} helps the eye catch
the key features of the situation involving two non-lightlike contours
$\mathcal{C}_1$ and $\mathcal{C}_2$.
For comparison, the smooth decomposition of a purely lightlike contour
is shown in the left panel.
In that case the junction point 3 creates no anomalous dimension and
the standard multiplication rule for gauge links applies.

In the above expression, $\Phi_{\rm cusp}$ contains a phase
entanglement ensuing from the renormalization effect on the cusp-like
junction point at infinity.
One may associate this phase with final (or initial) state
interactions, as proposed by Ji and Yuan in \cite{JY02}, and also
by Belitsky, Ji, and Yuan in \cite{BJY02}.
However, these authors (and also others) did not recognize that the
junction point in the split contour
(the latter stretching to light-cone infinity)
is no more a simple point, but a cusp obstruction that entails an
anomalous dimension  $\sim \ln p^+$.
More precisely, we have
\begin{equation}
\begin{split}
   & \gamma_{\rm cusp} (\alpha_s, \chi)
= \frac{\alpha_s}{\pi}C_{\rm F} \ (\chi \coth \chi - 1 ) \ , \\
& \frac{d}{d \ln p^+} \ \delta \gamma
= \lim_{\chi \to \infty}
  \frac{d}{d \chi} \gamma_{\rm cusp} (\alpha_s, \chi)
= \frac{\alpha_s}{\pi}C_{\rm F} \ ,
\label{eq:cusp-an-dim}
\end{split}
\end{equation}
which makes it apparent that the defect of the anomalous dimension is
related to the universal cusp anomalous dimension \cite{KR87}.
To derive this expression, we have used the fact that
$p^+ = (p \cdot n^-) \sim \cosh \chi$ defines an angle $\chi$
between the direction of the quark momentum $p_\mu$ and the lightlike
vector $n^-$.
Then, in the large $\chi$ limit, one has $\ln p^+ \to \chi$.
It is worth recalling in this context that the cusp anomalous
dimension of Wilson lines controls the Sudakov factor resulting from
gluon resummation and is known to the three-loop order \cite{VMV04}.
The Sudakov exponent in next-to-leading logarithmic approximation has
been calculated in \cite{SSK00} and expressed as an expansion in
inverse powers of the first beta-function coefficient.

\section{How to avert the defect of the anomalous dimension}
\label{sec:avert-an-dim}

\begin{figure}[t]
\centering
\includegraphics[scale=0.5,angle=0]{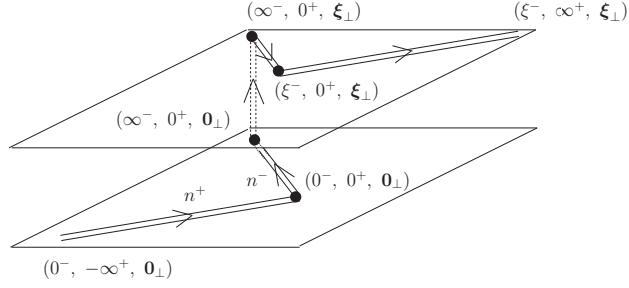}~~
\caption{Integration contour associated with the additional soft
         counter term.
\label{fig:contour}}
\end{figure}

In this section we will show in more depth how to get rid of the
cusp anomalous dimension and refurbish the definition of the TMD PDF.
The defect of the anomalous dimension, ensuing from the cusp-like
junction point of the non-lightlike gauge contours, represents a
distortion of the gauge-invariant formulation of the TMD PDF in the
light-cone gauge.
This is best appreciated by inspecting the composite non-smooth contour
$\mathcal{C}_{\rm cusp}$, visualized in Fig.\ \ref{fig:contour}, and
defined by
\begin{equation}
 \mathcal{C}_{\rm cusp}:\zeta_\mu
=
  \left\{
         [p_\mu^{+}s, - \infty < s < 0]
         \cup [n_\mu^-  s^{\prime},
         0 < s^{\prime} < \infty] \cup
         [ \mbox{\boldmath$l_\perp$} \tau , 0 < \tau < \infty ]
  \right\} \ ,
\label{eq:gpm}
\end{equation}
with $n_\mu^-$ being the minus light-cone vector.
This contour is obviously cusped: at the origin, the
four-velocity $p_\mu^{+}$, which is parallel to the plus
light-cone ray, is replaced---non-smoothly---by the four-velocity
$n_\mu^-$, which is parallel to the minus light-cone ray.
This means that exactly at this point the contour has a cusp, that is
characterized by the angle
$\chi \sim \ln p^+ = \ln (p \cdot n^-)$,
and will generate an anomalous dimension with the opposite sign
relative to $\delta\gamma$---cf.\ Eq.\ (\ref{eq:gamma_2}).
This contour can be used to define a soft counter term in the sense of
Collins and Hautmann \cite{CH00}, namely,
\begin{equation}
  R
\equiv
 \Phi (p^+, n^- | 0) \Phi^\dagger (p^+, n^- | \xi) \ ,
\label{eq:soft_factor_1}
\end{equation}
where the eikonal factors are given by
\begin{eqnarray}
\Phi (p^+, n^- | 0 )
 & = &
  \left\langle 0
  \left| {\cal P}
  \exp\Big[ig \int_{\mathcal{C}_{\rm cusp}}\! d\zeta^\mu
           \ t^a A^a_\mu (\zeta)
      \Big]
  \right|0
  \right\rangle \
\ , \\
  \Phi^\dagger (p^+, n^- | \xi )
 & = &
  \left\langle 0
  \left| {\cal P}
  \exp\Big[- ig \int_{\mathcal{C}_{\rm cusp}}\! d\zeta^\mu
           \ t^a A^a_\mu (\xi + \zeta)
      \Big]
  \right|0
  \right\rangle
\label{eq:soft_definition}
\end{eqnarray}
and have to be evaluated along the integration contour
$\mathcal{C}_{\rm cusp}$.

\begin{figure}[t]
\centering
\includegraphics[scale=0.65,angle=90]{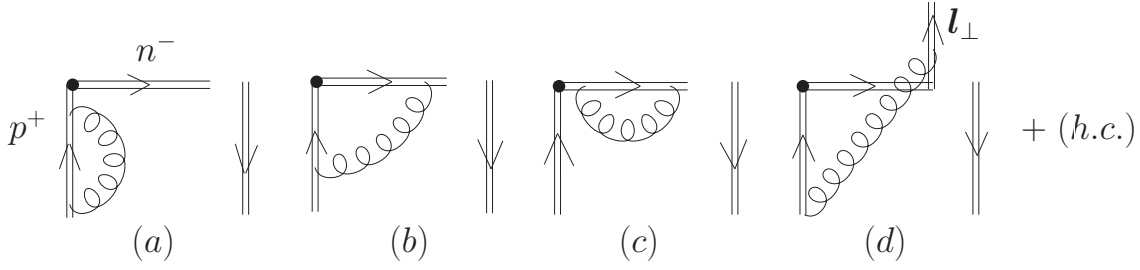}~~
\caption{Gluon radiative corrections giving rise to UV-divergences
         contributing to the soft counter term $R$.
         The designations are as in Fig.\ \ref{fig:se_gluon}.
\label{fig:soft_gluon}}
\end{figure}

Next, we consider the one-loop gluon radiative corrections,
contributing to the UV divergences of $R$ and displayed in
Fig.\ \ref{fig:soft_gluon}.
Diagrams (a) and (d) give rise to an anomalous dimension that
will finally compensate the anomalous-dimensions defect
generated by the cusp-like junction point of the contours.
On the other hand, by virtue of the light cone gauge $A^+=0$,
we are employing, diagrams (b) and (c) vanish.
The UV parts of diagrams (a) and (d) yield, respectively,
\begin{equation}
  \Phi^{(a)}_{\rm UV}(\eta)
=
  - \frac{\alpha_s}{\pi} C_{\rm F} \frac{1}{\epsilon}
  \left( \ln \frac{\eta}{p^+} - i\frac{\pi}{2} - i \pi C_\infty \right)
\label{eq:soft_a}
\end{equation}
and
\begin{equation}
  \Phi^{(d)}_{\rm UV}(\eta)
=
  - \alpha_s C_{\rm F} i \pi C_\infty \Gamma(\epsilon)
   \left(- 4\pi \frac{\mu^2}{\lambda^2}\right)^\epsilon \ .
\end{equation}
Combining these UV terms, we find
\begin{equation}
  F^{\rm (a + d)}_{\rm UV}(\eta)
=
  - \frac{\alpha_s}{\pi} C_{\rm F} \frac{1}{\epsilon}
  \left( \ln \frac{\eta}{p^+} - i\frac{\pi}{2}
    - i \pi C_\infty + i \pi C_\infty
  \right)
=
   - \frac{\alpha_s}{\pi}C_{\rm F}\frac{1}{\epsilon}
  \left( \ln \frac{\eta}{p^+} - i\frac{\pi}{2}
  \right) \ .
\end{equation}
Taking into account the Hermitian conjugate (``mirror'') terms, we
obtain the total UV-divergent part of the soft factor R in one-loop
order:
\begin{equation}
  \Phi^{\rm (1-loop)}_{\rm UV}(\eta)
  =
  - \frac{\alpha_s}{\pi}C_{\rm F}  \frac{2}{\epsilon}
  \ln \frac{\eta}{p^+} \ .
  \label{eq:phase_1loop}
\end{equation}
One notices that this expression bears no dependence on the pole
prescription, since all $C_\infty$-dependent terms have mutually
canceled.
Indeed, only the cusp-dependent term $\sim\ln \frac{\eta}{p^+}$
survives that will ultimately yield $-\gamma_{\rm cusp}$.

The above considerations make it apparent that one may use $R$ and
redefine the TMD PDF as follows
\begin{eqnarray}
f_{q/q}^{\rm mod}\left(x, \mbox{\boldmath$k_\perp$};\mu, \eta
                   \right)
\!\!\!\!\!\! && \!\!\! =
  \frac{1}{2}
  \int \frac{d\xi^- d^2\mbox{\boldmath$\xi_\perp$}}{2\pi (2\pi)^2}
  \exp\left(- i k^+ \xi^-
   + i \mbox{\boldmath$k_\perp$}
\cdot \mbox{\boldmath$\xi_\perp$}\right)
\Big\langle
              q(p) |\bar \psi (\xi^-, \mbox{\boldmath$\xi_\perp$})
\nonumber \\
&& \times
              [\xi^-, \mbox{\boldmath$\xi_\perp$};
   \infty^-, \mbox{\boldmath$\xi_\perp$}]^\dagger
   [\infty^-, \mbox{\boldmath$\xi_\perp$};
   \infty^-, \mbox{\boldmath$\infty_\perp$}]^\dagger
   \gamma^+[\infty^-, \mbox{\boldmath$\infty_\perp$};
   \infty^-, \mbox{\boldmath$0_\perp$}] \nonumber \\
&& \times
   [\infty^-, \mbox{\boldmath$0_\perp$}; 0^-,\mbox{\boldmath$0_\perp$}]
   \psi (0^-,\mbox{\boldmath$0_\perp$}) |q(p)
   \Big\rangle \nonumber \\
&& \times
   \Big[ \Phi(p^+, n^- | 0^-, \mbox{\boldmath$0_\perp$})
   \Phi^\dagger (p^+, n^- | \xi^-, \mbox{\boldmath$\xi_\perp$})
   \Big] \, .
\label{eq:final}
\end{eqnarray}
Before we conclude, let us mention that integrating the above
expression over the transverse momenta, we obtain an integrated
PDF that coincides with the standard one, containing no artifacts
of the cusped contour, and satisfying the DGLAP evolution equation.
Moreover,
$f_{q/q}^{\rm mod}\left(x, \mbox{\boldmath$k_\perp$};\mu, \eta
                   \right)
$
satisfies the simple renormalization-group equation
\begin{equation}
  \frac{1}{2} \mu \frac{d}{d\mu}
  \ln f_{q/q}^{\rm mod}(x, \mbox{\boldmath$k_\perp$}; \mu, \eta)
=
  \frac{3}{4} \frac{\alpha_s}{\pi}\, C_{\rm F} + O(\alpha_{s}^{2})\ .
\end{equation}
Note that without the soft counter term, $R$, extra contributions
to the anomalous dimension on the right-hand side would appear.
In \cite{CS08} we have outlined the correspondence between
the evolution with respect to the scale parameter $\eta$ in our approach
and the Collins-Soper evolution equation with respect to the rapidity
parameter $\zeta$, establishing the absence of UV singularities
entailed by the light-cone gauge.

\section{Conclusions}
\label{sec:concl}

To summarize the results of this report on the renormalization of
gauge-invariant TMD PDFs, the following may be said.
First, we have elaborately discussed the one-loop calculation of the
UV divergences of a typical TMD PDF which contains lightlike and
transverse gauge links in order to fully restore gauge invariance.
We found that an extra UV divergence appears, not noticed before in
the literature, which is unrelated to the quark self energy and the
end-point singularities of the contours.
Second, we showed that these divergences give rise to an anomalous
dimension, which can be regarded as originating from the
renormalization effect on a cusp-like junction point of the integration
contours in the gauge links at light-cone infinity.
At the considered one-loop order, this anomalous dimension coincides
with the universal cusp anomalous dimension of Wilson-line operators
and is an ingrained property of the split contours.
Third, in order to dispense with this anomalous-dimensions defect
and recover the well-known results in a covariant gauge (say, in the
Feynman gauge) in which $\mathbf{A}_{\perp}$ vanishes at infinity,
we have proposed a modified definition of the TMD PDF.
This definition includes a Collins-Hautmann soft counter term by means
of path-ordered eikonal factors that are evaluated along a specific
non-smooth contour off the light cone.
This cusped contour suffices to neutralize the cusp artifact
encountered in the standard definition of the TMD PDF.
Finally, as we outlined in \cite{CS08}, the soft counter term can be
given an interpretation akin to the ``intrinsic'' Coulomb phase found
by Jakob and Stefanis \cite{JS90} in QED.
In both cases, a phase entanglement appears, ensuing either from the
charged ``particle behind the moon'' (QED) or from the cusp-like
junction point at light-cone infinity (QCD).
Recently, Collins \cite{Col08} has considered possible refinements and
modifications in the definition of unintegrated parton densities that
deserve further examination.
An improved definition of TMD PDFs will have tangible consequences in
several areas of QCD.


\end{document}